%% file: ms.tex
\documentclass[letterpaper, 10 pt, conference]{ieeeconf} 
\IEEEoverridecommandlockouts   

\usepackage{cite}
\usepackage{amsmath,amssymb,amsfonts}
\usepackage{xcolor}
\def\BibTeX{{\rm B\kern-.05em{\sc i\kern-.025em b}\kern-.08em T\kern-.1667em\lower.7ex\hbox{E}\kern-.125emX}}

\usepackage{amsthm}
\newtheoremstyle{dotless}{}{}{\itshape}{}{\bfseries}{}{ }{}

\let\labelindent\relax 
\usepackage{enumitem}

\usepackage{graphicx}
\usepackage{placeins}
\usepackage{pgfplots}
\pgfplotsset{compat=newest}
\usepackage{xurl}
\usepackage{eso-pic}

\usepackage{etoolbox}
\newbool{long}
\setbool{long}{false}

\newcommand{\inputnewtext}[1]{\ifbool{long}{\textcolor{blue}{#1}}{#1}}
\newcommand{\deleteoldtext}[1]{\ifbool{long}{\textcolor{red}{#1}}{}}

\theoremstyle{dotless}
\newtheorem{dummy}{}

\newtheorem{theorem}[dummy]{Theorem}
\newtheorem{corollary}[dummy]{Corollary}
\newtheorem{lemma}[dummy]{Lemma}

\newtheorem{definition}[dummy]{Definition}
\newtheorem{remark}[dummy]{Remark}

\newcommand{\norm}[1]{\lVert#1\rVert}

\newcommand{\T}{\top\!}

\newcommand{\vecc}{\textnormal{vec}}

\input{my_colors.tex}

\newcommand{\drawlinelegend}[1]{\raisebox{.5ex}{\tikz{\draw[#1, line width=0.4mm] (0,0) -- +(1em, 0);}}}
\newcommand{\drawrectanglelegend}[2]{\raisebox{.0ex}{\tikz{\filldraw[color=#1!100, fill=#1!#2] (0,0) rectangle (2ex, 1ex);}}}
\newcommand{\drawrectanglelegendsmall}[2]{\raisebox{.25ex}{\tikz{\filldraw[color=#1!100, fill=#1!#2] (0,0) rectangle (2ex, .5ex);}}}

\begin{document}

\AddToShipoutPictureBG*{%
	\AtPageUpperLeft{%
		\setlength\unitlength{1in}%
		\hspace*{\dimexpr0.5\paperwidth\relax}
		\makebox(0,-0.5)[c]{\begin{tabular}{c c}
				Johan Kon \textit{et al}, 	"Unconstrained Parameterization of Stable LPV Input-Output Models: with Application to System Identification", \\
				accepted for publication in the Proceedings of the European Control Conference, 2024. Uploaded \today \\
		\end{tabular}}
}}

\bstctlcite{IEEEexample:BSTcontrol} 

\title{\LARGE \bf Unconstrained Parameterization of Stable LPV Input-Output Models: \\
with Application to System Identification}

\author{Johan Kon$^1$, Jeroen van de Wijdeven$^2$, Dennis Bruijnen$^3$, Roland T\'{o}th$^{4}$, Marcel Heertjes$^{1,2}$, Tom Oomen$^{1,5}$ 
\thanks{This work is supported by Topconsortia voor Kennis en Innovatie (TKI), and ASML and Philips Engineering Solutions. $^1$: Control Systems Technology Group, Departement of Mechanical Engineering, Eindhoven University of Technology, Eindhoven, The Netherlands, e-mail: j.j.kon@tue.nl. $^2$: ASML, Veldhoven, The Netherlands. $^3$: Philips Engineering Solutions, Eindhoven, The Netherlands. $^4$: Control Systems Group, Electrical Engineering, Eindhoven University of Technology, The Netherlands, and the HUN-REN Institute for Computer Science and Control, Budapest, Hungary. $^5$: Delft University of Technology, Delft, The Netherlands. }
}

\maketitle

\input{./Sections/Abstract.tex}

\input{./Sections/Introduction.tex}

\input{./Sections/Problem_formulation.tex}

\input{./Sections/Contracting_LPV_IO_models.tex}

\input{./Sections/Convex_and_direct_parametrization_of_contracting_LPV_IO_models.tex}

\input{./Sections/Simulation_example.tex}

\input{./Sections/Conclusion.tex}

\bibliographystyle{IEEEtran}
\bibliography{IEEEabrv,BSTcontrol.bib,library.bib}

\end{document}

%% file: my_colors.tex
\definecolor{red}{rgb}{1,0,0}
\definecolor{blue}{rgb}{0,0,1}
\definecolor{green}{rgb}{0,1,0}
\definecolor{yellow}{rgb}{1,1,0}
\definecolor{orange}{rgb}{1,0.647,0}
\definecolor{gold}{rgb}{1,0.843,0}
\definecolor{purple}{rgb}{0.627,0.125,0.941}
\definecolor{gray}{rgb}{0.745,0.745,0.745}
\definecolor{brown}{rgb}{0.647,0.165,0.165}
\definecolor{navy}{rgb}{0,0,0.502}
\definecolor{pink}{rgb}{1,0.753,0.796}
\definecolor{seagreen}{rgb}{0.18,0.545,0.341}
\definecolor{turquoise}{rgb}{0.251,0.878,0.816}
\definecolor{violet}{rgb}{0.933,0.51,0.933}
\definecolor{darkblue}{rgb}{0,0,0.545}
\definecolor{darkcyan}{rgb}{0,0.545,0.545}
\definecolor{darkgray}{rgb}{0.663,0.663,0.663}
\definecolor{darkgreen}{rgb}{0,0.392,0}
\definecolor{darkmagenta}{rgb}{0.545,0,0.545}
\definecolor{darkorange}{rgb}{1,0.549,0}
\definecolor{darkred}{rgb}{0.545,0,0}
\definecolor{lightblue}{rgb}{0.678,0.847,0.902}
\definecolor{lightcyan}{rgb}{0.878,1,1}
\definecolor{lightgray}{rgb}{0.827,0.827,0.827}
\definecolor{lightgreen}{rgb}{0.565,0.933,0.565}
\definecolor{lightyellow}{rgb}{1,1,0.878}
\definecolor{black}{rgb}{0,0,0}
\definecolor{white}{rgb}{1,1,1}
\definecolor{mblue}{rgb}{0, 0.4470, 0.7410}
\definecolor{mred}{rgb}{0.85, 0.325, 0.098}
\definecolor{morange}{rgb}{0.929, 0.694, 0.125}
\definecolor{mpurple}{rgb}{0.494, 0.184, 0.556}
\definecolor{mgreen}{rgb}{0.466, 0.674, 0.188}
\definecolor{mcyan}{rgb}{0.301, 0.745, 0.933}
\definecolor{mbrown}{rgb}{0.635, 0.078, 0.184}

%% file: Sections/Abstract.tex
\begin{abstract}
Ensuring stability of discrete-time (DT) linear parameter-varying (LPV) input-output (IO) models estimated via system identification methods is a challenging problem as known stability constraints can only be numerically verified, e.g., through solving Linear Matrix Inequalities. 
In this paper, an unconstrained DT-LPV-IO parameterization is developed which gives a stable model for any choice of model parameters. 
To achieve this, it is shown that \textit{all} quadratically stable DT-LPV-IO models can be generated by a mapping of transformed coefficient functions that are constrained to the unit ball, i.e., a small-gain condition. The unit ball is then reparameterized through a Cayley transformation, resulting in an unconstrained parameterization of all quadratically stable DT-LPV-IO models. 
As a special case, an unconstrained parameterization of all stable DT linear time-invariant transfer functions is obtained. 
Identification using the stable DT-LPV-IO model with neural network coefficient functions is demonstrated on a simulation example of a position-varying mass-damper-spring system.
\end{abstract}

%% file: Sections/Introduction.tex
\section{Introduction}
\label{sec:introduction}
Ever stringent performance requirements from practice necessitate \deleteoldtext{the need} to also model and identify the nonlinear behavior of systems \cite{Schoukens2019}. These nonlinear characteristics make modeling these systems based on first-principles increasingly difficult and time-consuming, such that it becomes necessary to adopt data-driven system modeling tools, i.e., system identification.

Linear parameter-varying (LPV) systems \cite{Toth2010} are a powerful surrogate system class for capturing nonlinear and time-varying behaviour. In LPV systems, the signal relations are linear, but the coefficients describing these relations are a function of a time-varying scheduling signal $\rho$ that is assumed to be measurable online. The resulting parameter-varying behavior can embed certain nonlinear characteristics under the correct choice of $\rho$ \cite{Leith2000a}. 
Data-driven system identification methods for LPV systems have been thoroughly developed in the last decades, both for input-output (IO) \cite{Bamieh2002,Laurain2010,Zhao2012a} as well as state-space (SS) representations \cite{Felici2007,Cox2018a}. 

Even though many systems to be modeled in an LPV form are known to be stable, ensuring stability of the identified model is a challenging problem for two reasons. First, the parameter estimates are sensitive to modeling errors, finite data effects, and measurement noise, such that the identified model can be unstable even if the underlying system is stable \cite{Lacy2003,Haene2006}. Second and most importantly, for LPV models there is no analytic explicit constraint on the model parameters that ensures stability. Instead, stability \deleteoldtext{of the model} can only be verified \deleteoldtext{numerically}\inputnewtext{implicitly}, e.g., by testing the feasibility of Linear Matrix Inequalities (LMIs) representing a stability condition \cite{Apkarian1995, Scherer2001}.

To ensure stability of the identified model, recently unconstrained \textit{state-space} (SS) models have been developed that are stable for any choice of model parameters \cite{Revay2023_REN, Martinelli2023, Verhoek_CDC_2023}. This is achieved 
by reparameterizing the stability constraint in the form of an LMI through new unconstrained parameters in a necessary and sufficient manner, such for any choice in these new parameters, the LMI is satisfied, i.e., stability is guaranteed.
In \cite{Revay2023_REN}, such an unconstrained parameterization is developed for discrete-time (DT) Lur'e models with neural network nonlinearities. \inputnewtext{This technique has subsequently been applied to continuous-time Lur'e models \cite{Martinelli2023} and discrete-time LPV SS models \cite{Verhoek_CDC_2023}.}
\deleteoldtext{The continuous-time case is subsequently developed in \cite{Martinelli2023}, and in \cite{Verhoek_CDC_2023} an unconstrained parameterization is derived for the LPV-SS case.}

In contrast, to guarantee stability of an identified \textit{LPV input-output} (LPV-IO) model, current methods either explicitly enforce a stability constraint during identification \cite{Cerone2012, Cerone2012a,Wollnack2016,Wollnack2017} or restrict the model class such that it is a priori guaranteed to be stable \cite{Henrion2003,Gilbert2007}. However, enforcing a stability constraint - usually in the form of an LMI or sum of squares condition - during identification severely increasing the computational complexity of the optimization \cite{Lacy2003, Revay2021}. Alternatively, restricting the model class is usually done according to a simple but often conservative approximation of all stable models, limiting the representation capability \deleteoldtext{of the model}.

To overcome the computational complexity and conservatism of the previous approaches, in this paper it is shown that \textit{all} quadratically stable (QS) DT-LPV-IO models can be generated by a mapping of unconstrained transformed coefficient functions. 
The approach is based on reparameterizating the LPV coefficient functions such that the LMI representing the QS condition is satisfied for any choice of model parameters, similar to the state-space case \cite{Revay2023_REN, Revay2022, Martinelli2023, Verhoek_CDC_2023}. The difficulty in obtaining unconstrained stable LPV-IO models is that a direct state construction for IO models is based on the inputs and outputs and their past values. Consequently, the equations describing the state evolution over time are structured, such that not all components of the resulting LMI stability condition can be freely chosen. This is in contrast to the SS case in which the state is not fixed, resulting in an unstructured problem.
\inputnewtext{A consequence of this reparameterization is that still only QS systems can be represented. Thus, an LPV system that is stable but not quadratically cannot be exactly represented. However, quadratic stability already represents a significant improvement over the current conservative approximations, and constrained identification approaches also enforce QS.}

The main contribution of this paper is an unconstrained parameterization of all quadratically stable DT-LPV-IO models, allowing for unconstrained system identification with a priori stability guarantees. The set of stable DT transfer functions is obtained as a special case. The main contribution consists of the following sub-contributions.
\begin{enumerate}[label=C\arabic*)]
    \item A criterion in the form of a matrix inequality to test stability of LPV-IO models (Section \ref{sec:contracting_LPV_IO_models}), and a corresponding graphical interpretation of the allowed coefficient function value sets in which stability is guaranteed (Section \ref{sec:simulation_example}). 
    \item A reparameterization of the coefficient functions of LPV-IO models such that above criterion is satisfied for any choice of the new model parameters (Section \ref{sec:unconstrained_contraction_LPV_IO}).
    \item A simulation example demonstrating the applicability of the developed method (Section \ref{sec:simulation_example}).
\end{enumerate}

\subsubsection*{Notation} $\norm{\cdot}_2$ represents the Euclidean vector norm. $\mathbb{Z}_{\geq 0}$ represents the set of non-negative integers. 
A \textit{symmetric} matrix $P \in \mathbb{R}^{n \times n}$ is said to be positive definite if $x^\T P x > 0 \ \forall x \in \mathbb{R}^n \setminus \{ 0\}$, also denoted by $P \succ 0$ and $P \in \mathbb{S}_{\succ 0}$. Similarly, $P\prec 0$ and $P \in \mathbb{S}_{\prec 0}^n$ denote negative-definiteness. 

%% file: Sections/Problem_formulation.tex
\section{Problem Formulation}
\label{sec:problem_formulation}
Consider the LPV-IO model with input $u_k \in \mathbb{R}$ and output $y_k \in \mathbb{R}$ represented by the difference equation
\begin{equation}
    \label{eq:LPV_IO}
    y_k= -\sum_{i=1}^{n_a} a_i(\rho_k) y_{k-i} + \sum_{i=0}^{n_b-1} b_i(\rho_k) u_{k-i},
\end{equation}
with coefficient functions $a_i(\rho),b_i(\rho): \mathbb{P} \rightarrow \mathbb{R}$ describing the dependence of the difference equation on the scheduling signal $\rho_k \in \mathbb{P} \subseteq \mathbb{R}^{n_\rho}$, time index $k \in \mathbb{Z}_{\geq 0}$, and model order $n_a \geq 0$, $n_b \geq 1$.

The coefficient functions are parameterized by a function $g_\phi(\rho)$ depending on model parameters $\phi \in \mathbb{R}^{n_\phi}$, i.e.,
\begin{equation}
\label{eq:coefficient_functions}
g_\phi(\rho) = \begin{bmatrix} a_{1}(\rho) & a_2(\rho)&  \cdots & b_{n_b-1}(\rho)\end{bmatrix}^\T.
\end{equation}
Examples include affine coefficient functions, e.g., $g_\phi^\text{aff}(\rho) \! = \! E \rho \! + \! c$ with $\phi \! = \!\vecc(E,c)$, polynomial basis function expansions \cite{Zhao2012a}, e.g., $g_\phi^\text{pol}(\rho) = c + E_1 \rho + E_2 \rho^2 + \ldots + E_{d} \rho^d$ with $\phi = \vecc(E_1,\ldots,E_d,c)$, or a neural network \cite{Previdi2003, KonCDC2023}
\begin{equation*}
    g_{\phi}^\text{NN} = E_\mathrm{L} \sigma( E_{\mathrm{L}-1} \sigma( \cdots ( E_0 \rho + c_0) \cdots) + c_{\mathrm{L}-1}) + c_L,
    \label{eq:NN}
\end{equation*}
with nonlinear activation function $\sigma$, e.g., $\sigma(\cdot) = \tanh(\cdot)$, and parameters $\phi = \vecc(E_0,c_0,\ldots E_\mathrm{L},c_\mathrm{L})$.
\begin{remark}
    For ease of notation, $u_k,y_k \in \mathbb{R}$ is considered. However, all results immediately extend to the setting in which $u_k \in \mathbb{R}^{n_u}$, $y_k \in \mathbb{R}^{n_y}$.
\end{remark}

Given model class \eqref{eq:LPV_IO} and a parameterization of $g_\phi(\rho)$, a natural question is if the model is stable for the chosen model parameters $\phi$ and the relevant range of $\rho$.
Here, stability is defined as the model output $y_k$ asymptotically going to zero for zero input, as defined next.
\begin{definition}
    Given a parameterization $g_\phi: \mathbb{P} \rightarrow \mathbb{R}^{n_a + n_b}$ with model parameters $\phi \in \mathbb{R}^{n_\phi}$, the LPV-IO model \eqref{eq:LPV_IO} is said to be uniformly asymptotically stable if for any time $\bar{k}$, any scheduling signal $\rho$ with $\rho_k \in \mathbb{P}$, and any input $u$ with $u_k = 0 \ \forall k > \bar{k}$, the response $y_k$ of \eqref{eq:LPV_IO} satisfies $\lim_{k \rightarrow \infty} y_k = 0$.
    \label{def:stability}
\end{definition}
Note that by standard dissipativity arguments, stability as in Definition \ref{def:stability} also implies that the LPV-IO model has finite induced $\ell_2$ gain, i.e., $\exists \gamma < \infty$ such that $\sum_{k=0}^\infty \norm{y_k}_2^2< \gamma^2 \sum_{k=0}^\infty \norm{u_k}_2^2$, see, e.g., \cite{Revay2023_REN}. Additionally, by linearity of \eqref{eq:LPV_IO}, also incremental properties follow.

The goal of this paper then is to parameterize $g_\phi(\rho)$ in \eqref{eq:coefficient_functions} in such a way that LPV-IO model \eqref{eq:LPV_IO} is stable for any choice of $\phi$, i.e., guaranteeing stability without constraints, for any parameterization of $g_\phi$ (e.g., NN). Of course for stability such a $g_\phi$ should necessarily result in finite $a_i(\rho),b_i(\rho)$ for all possible $\rho$, i.e., $\norm{g_\phi(\rho)}_2 < \infty \ \forall \rho \in \mathbb{P}$, which is taken as a precondition in the remainder of the paper.


%% file: Sections/Contracting_LPV_IO_models.tex
\section{Stable LPV-IO models}
\label{sec:contracting_LPV_IO_models}
To obtain an unconstrained parameterization of \eqref{eq:LPV_IO}, first a condition for determining stability of \eqref{eq:LPV_IO} as in Defintion \ref{def:stability} in terms of $a_i(\rho),b_i(\rho)$ is required. This section derives such a stability condition in the form of a matrix inequality, constituting contribution C1.

\subsection{Maximum State-space Representation}
To derive a stability condition, first it is required to define a state for \eqref{eq:LPV_IO} such that standard Lyapunov techniques can be used. To avoid issues around minimum realizations in absence of any structure in the coefficient functions \cite{Toth2012b}, a non-minimum state-space representation for \eqref{eq:LPV_IO} is used \cite{Toth2013}. 

More specifically, \eqref{eq:LPV_IO} can be equivalently represented as \eqref{eq:maximum_state_space} with state $x$ storing the previous inputs and outputs as
\begin{equation}
\begin{aligned}
    x_k &= \begin{bmatrix} \bar{y}_k^\T & \bar{u}_k^\T \end{bmatrix}^\T \in \mathbb{R}^{n_a+ n_b} \\
    \bar{y}_k &= \begin{bmatrix} y_{k-1} & \hdots & y_{k-n_a} \end{bmatrix}^\T \in \mathbb{R}^{n_a} \\
    \bar{u}_k &= \begin{bmatrix} u_{k-1} & \hdots & u_{k-n_b+1} \end{bmatrix}^\T \in \mathbb{R}^{n_b-1}.
\end{aligned}
\label{eq:maximum_ss_state_def}
\end{equation}
\begin{figure*}[t]
\begin{subequations}
\label{eq:maximum_state_space}
\begin{equation}
\label{eq:maximum_state_space_state}
\scalebox{.83}{$
    x_{k+1}
    = 
    \left[
    \begin{array}{ccccc|ccccc}
        -a_1(\rho_k) & -a_2(\rho_k) & \hdots & -a_{n_a-1}(\rho_k) & -a_{n_a}(\rho_k) & b_1(\rho_k) & b_2(\rho_k) & \hdots & b_{n_b-2}(\rho_k) & b_{n_b-1}(\rho_k) \\
        1 &  & & \emptyset & 0 & 0 & 0 & \hdots & 0 & 0\\
         & 1 &  &  & 0 & 0 & 0 & \hdots & 0 & 0\\
        & & \ddots  &  & \vdots & \vdots & \vdots & & \vdots & \vdots \\
        \emptyset &  & & 1 & 0 & 0 & 0 & \hdots & 0 & 0\\
        \hline
        0 & 0 & \hdots & 0 & 0 & 0 & 0 & \hdots & 0 & 0\\
        0 & 0 & \hdots & 0 & 0 & 1 &  &  & \emptyset & 0\\
        0 & 0 & \hdots & 0 & 0 &  & 1 &  &  & 0\\
        \vdots & \vdots & & \vdots & \vdots & & & \ddots & & \vdots \\
        0 & 0 & \hdots & 0 & 0 & \emptyset & & & 1 & 0
    \end{array}
    \right]
    x_k
    + 
    \left[
    \begin{array}{c}
        b_0(\rho_k) \\
        0 \\
        0 \\
        \vdots \\
        0 \\
        \hline
        1 \\
        0 \\
        0 \\
        \vdots \\
        0
    \end{array}
    \right]
    u_k$}
\end{equation}
\begin{equation} 
\label{eq:maximum_state_space_output}
\scalebox{.84}{$
    \hspace{-35pt} y_k = \left[ \begin{array}{ccccc|ccccc} 
        -a_1(\rho_k) & -a_2(\rho_k) & \hdots & -a_{n_a-1}(\rho_k) & -a_{n_a}(\rho_k) & b_1(\rho_k) & b_2(\rho_k) & \hdots&  b_{n_b-2}(\rho_k) & b_{n_b-1}(\rho_k)
    \end{array}\right] x_k
    + b_0(\rho_k) u_k
    $}
\end{equation}
\end{subequations}
\end{figure*}
Representation \eqref{eq:maximum_state_space} is compactly written as
\begin{equation}
    \label{eq:maximum_ss_short}
    \begin{aligned}
        x_{k+1} &= A(\rho_k) x_k + B(\rho_k) u_k \\
        y_k &= C(\rho_k) x_k + D(\rho_k) u_k,
    \end{aligned}
\end{equation}
where
\begin{align}
    A(\rho) &= 
    \left[\begin{array}{c|c}
        F - G K(\rho) & G L(\rho) \\
        \hline \hspace{-5pt} 
        0 & F_b
    \end{array}\right] 
    &
    B(\rho) &= \left[\begin{array}{c}
    G b_0 (\rho) \\
    \hline
    G_b
    \end{array}\right] \nonumber \\
    \label{eq:maximum_ss_ABCD}
    C(\rho) &= \left[\begin{array}{c|c}
         -K(\rho) & L(\rho)
    \end{array}\right] & D(\rho) &= b_0(\rho).
\end{align}
Matrices $F,F_b$ and $G,G_b$ correspond to a discrete-time buffer system that collects past samples of $y_k$ and $u_k$, i.e.,
\begin{align}
    F &= \begin{bmatrix}
        0 & 0 \\
        I_{n_a-1} & 0
    \end{bmatrix} \in \mathbb{R}^{n_a \times n_a}
    &  &
    G = \begin{bmatrix}
        1 \\ 0
    \end{bmatrix}
    \in \mathbb{R}^{n_a}, \label{eq:maximum_ss_buffer}
    \\
    F_b &= \begin{bmatrix}
        0 & 0 \\
        I_{n_b-2} & 0
    \end{bmatrix} \in \mathbb{R}^{n_b -1 \times n_b -1}
    & &
    G_b = \begin{bmatrix}
        1 \\ 0
    \end{bmatrix}
    \in \mathbb{R}^{n_b-1}, \nonumber
\end{align}
with $0$ entries of appropriate dimensions. $K(\rho), L(\rho)$ collect coefficient functions $a_i(\rho),b_i(\rho)$ as
\begin{equation}
\begin{aligned}
    K(\rho) &= \begin{bmatrix} a_1(\rho) &  \hdots & a_{n_a-1}(\rho) & a_{n_a}(\rho) \end{bmatrix} \in \mathbb{R}^{n_a} \\
    L(\rho) &= \begin{bmatrix} b_1(\rho) &  \hdots & b_{n_b-2}(\rho) & b_{n_b-1}(\rho) \end{bmatrix} \in \mathbb{R}^{n_b-1}. \label{eq:maximum_ss_L_K}  
\end{aligned}
\end{equation}
Although representation \eqref{eq:maximum_state_space} is not a minimal representation of \eqref{eq:LPV_IO}, if \eqref{eq:LPV_IO} is controllable and observable, \eqref{eq:maximum_state_space} is controllable and detectable, 
hence it can be used to directly characterize stability of \eqref{eq:LPV_IO}.

\subsection{Quadratically Stable LPV-IO Models}
Representation \eqref{eq:maximum_ss_short} allows for using standard Lyapunov techniques to determine stability. Specifically, stability can be analyzed by considering a common quadratic Lyapunov function, as formalized next.

\begin{lemma}
    \label{lem:quadratic_stability}
    Given parameters $\phi \in \mathbb{R}^{n_\phi}$, the LPV-IO model \eqref{eq:LPV_IO} is stable as in Definition \ref{def:stability} if there exists a $\mathcal{P} \succ 0$ such that
    \begin{equation}
        \mathcal{P} - A^\T(\rho) \mathcal{P} A(\rho) \succ 0 \quad \forall \rho \in \mathbb{P}.
        \label{eq:quadratic_stability}
    \end{equation}
\end{lemma}
\begin{proof}
    The proof follows by standard Lyapunov arguments. For any initial condition $x_{\bar{k}}$ at time $\bar{k}$ with $x$ as in \eqref{eq:maximum_ss_state_def} and input $u_k = 0 \ \forall k > \bar{k}$, the state of \eqref{eq:maximum_state_space_state} evolves as $x_{k+1} = A(\rho) x_k$. Next define $V(x) = x^\T \mathcal{P} x$ such that along trajectories of \eqref{eq:maximum_state_space_state}, $V(x)$ satisfies
    \begin{equation}
        V(x_{k+1}) - V(x_k) = x_k^\T ( A^\T(\rho_k) \mathcal{P} A(\rho_k) - \mathcal{P}) x_k.
    \end{equation} 
    Thus, if $\mathcal{P} \succ 0$ and it satisfies \eqref{eq:quadratic_stability}, $V(x)$ is a positive-definite function that is decreasing along trajectories of \eqref{eq:maximum_state_space}, such that $\lim_{k\rightarrow\infty} V(x_k) = 0$, implying that $\lim_{k\rightarrow\infty} x_k = 0$ and thus $\lim_{k\rightarrow\infty} y_k = 0$, completing the proof.
\end{proof}

All LPV-IO models that satisfy Lemma \ref{lem:quadratic_stability} are called quadratically stable (QS). Condition \eqref{eq:quadratic_stability} provides a computational test for determining QS of the LPV-IO model class \eqref{eq:LPV_IO}, showing that determining stability of IO parameterizations can be addressed using state-space methods.

%% file: Sections/Convex_and_direct_parametrization_of_contracting_LPV_IO_models.tex
\section{Unconstrained Parameterization of Stable LPV-IO models}
\label{sec:unconstrained_contraction_LPV_IO}
In this section, it is shown that there exists $\mathcal{P},K(\rho),L(\rho)$ that satisfy \eqref{eq:quadratic_stability} if and only if there exist unconstrained variables $X(\rho),Z(\rho),W$ related to $\mathcal{P},K(\rho)$ in a one-to-one way and that $L(\rho)$ is irrelevant for satisfying \eqref{eq:quadratic_stability}. These relations then allow for an unconstrained reparameterization of the coefficient functions such that \eqref{eq:quadratic_stability} is always satisfied, constituting Contribution C2. Consequently, any choice of model parameters $\phi$ of this unconstrained reparameterization results in coefficient functions for which \eqref{eq:LPV_IO} is stable.

\subsection{Eliminating the Influence of $L(\rho)$}
First it is shown that only $F - G K(\rho)$ has to be considered to satisfy the stability condition \eqref{eq:quadratic_stability}, i.e., $L(\rho)$ is already unconstrained. Intuitively, since $A(\rho)$ is upper block diagonal, only its diagonal blocks $F-G K(\rho)$ and $F_b$ determine stability. However, $F_b$ represents a simple LTI chain of delays and is trivially stable, meaning that the stability of $F-GK(\rho)$ determines the stability of \eqref{eq:LPV_IO}. These claims are formalized by the following lemma.
\begin{lemma}
\label{lem:contracting_series_interconnection}
    For a given $A(\rho)$ as in \eqref{eq:maximum_ss_ABCD}, there exists a $\mathcal{P} \succ 0$ satisfying \eqref{eq:quadratic_stability} if and only if there exists a $P \succ 0$ such that
    \begin{align}
        P - (F - G K(\rho))^\T P (F - G K(\rho)) &\succ 0 \quad \forall \rho \in \mathbb{P} \label{eq:contraction_F_GK},
    \end{align} and $\norm{L(\rho)}_2 < \infty \ \forall \rho \in \mathbb{P}$.
\end{lemma}
\begin{proof}
Define $\mathcal{P} = \begin{bmatrix} \mathcal{P}_{11} & \mathcal{P}_{12} \\ \mathcal{P}_{12}^\T & \mathcal{P}_{22}\end{bmatrix}$. Substitute in \eqref{eq:quadratic_stability} to obtain $\mathcal{P} - A^\T(\rho) \mathcal{P} A(\rho) = \mathcal{Q}(\rho)=\begin{bmatrix} \mathcal{Q}_{11}(\rho) & \mathcal{Q}_{12}(\rho) \\ \mathcal{Q}_{12}^\T(\rho) & \mathcal{Q}_{22}(\rho) \end{bmatrix} $
with
\begin{equation*}
\begin{aligned}
    \mathcal{Q}_{11}(\rho) =& -(F-GK(\rho))^\T \mathcal{P}_{11} (F-GK(\rho)) + \mathcal{P}_{11} \\
    \mathcal{Q}_{12}(\rho) =& -(F - GK(\rho))^\T (\mathcal{P}_{11} G L(\rho) + \mathcal{P}_{12} F_b) + \mathcal{P}_{12 } \\
    \mathcal{Q}_{22}(\rho) =& -L^\T(\rho) G^\T \mathcal{P}_{11} G L(\rho) - L^\T(\rho) G^\T \mathcal{P}_{12} F_b \\
    &- F_b^\T \mathcal{P}_{12} G L(\rho) - F_b^\T \mathcal{P}_{22} F_b + \mathcal{P}_{22}.
\end{aligned}
\end{equation*}
With these relations, Lemma \ref{lem:contracting_series_interconnection} is proven as follows. Throughout, $\mathcal{Q} \succ_{\mathbb{P}} 0$ is shorthand for $\mathcal{Q}(\rho) \succ 0 \ \forall \rho \in \mathbb{P}$.

\eqref{eq:quadratic_stability} $\Rightarrow$ \eqref{eq:contraction_F_GK}: If \eqref{eq:quadratic_stability} is satisfied, then $\mathcal{Q} \succ_{\mathbb{P}} 0$, and thus $\mathcal{Q}_{11} \succ_{\mathbb{P}} 0$, such that \eqref{eq:contraction_F_GK} is satisfied with $P = \mathcal{P}_{11}$.

\eqref{eq:quadratic_stability} $\Leftarrow$ \eqref{eq:contraction_F_GK}: $\mathcal{P} - A^\T(\rho) \mathcal{P} A(\rho) = \mathcal{Q} \succ_{\mathbb{P}} 0$ iff $\mathcal{Q}_{11} \succ_{\mathbb{P}} 0$ and its Schur complement $\mathcal{Q}_{22} - \mathcal{Q}_{12}^\T \mathcal{Q}_{11}^{-1} \mathcal{Q}_{12} \succ_{\mathbb{P}} 0$. Now, since \eqref{eq:contraction_F_GK} holds, set $\mathcal{P}_{11} = P$ to obtain $\mathcal{Q}_{11} \succ_{\mathbb{P}} 0$. Additionally, since $F_b$ is stable, there exists a $\mathcal{P}_{22} \succ 0$ such that the term $- F_b^\T \mathcal{P}_{22} F_b + \mathcal{P}_{22}$ in $\mathcal{Q}_{22}$ can be made arbitrary large, i.e., large enough to ensure that also $\mathcal{Q}_{22} - \mathcal{Q}_{12}^\T \mathcal{Q}_{11}^{-1} \mathcal{Q}_{12} \succ_{\mathbb{P}} 0$, such that $\mathcal{Q} \succ_{\mathbb{P}} 0$. 
Specifically, set $\mathcal{P}_{12} =0$ and define $\kappa = \max_{\rho \in \mathbb{P}} \norm{\mathcal{Q}_{12}(\rho)}_2$, $\lambda = \max_{\rho \in \mathbb{P}} \norm{L^\T(\rho) G^\T \mathcal{P}_{11} G L(\rho)}_2 $, and $\mu = \max_{\rho \in \mathbb{P}} \norm{\mathcal{Q}_{11}^{-1}(\rho)}_2$, which are all well-defined since $K(\rho)$ and $L(\rho)$ are bounded $\forall \rho \in \mathbb{P}$ as they are composed from $a_i(\rho)$ and $b_i(\rho)$ which are all bounded on $\mathbb{P}$. Then any $\mathcal{P}_{22}$ such that $- F_b^\T \mathcal{P}_{22} F_b + \mathcal{P}_{22} \succ (\lambda + \kappa^2/\mu)I$ holds, ensures that $\mathcal{Q} \succ_{\mathbb{P}} 0$, implying \eqref{eq:quadratic_stability}. Lastly, $\mathcal{P}_{11},\mathcal{P}_{22} \succ 0$ and $\mathcal{P}_{12}=0$ guarantee $\mathcal{P} \succ 0$, completing the proof.
\end{proof}
Consequently, in the remainder, only \eqref{eq:contraction_F_GK} is required to be considered, parameterizing all quadratic Lyapunov functions for $F - G K(\rho)$ corresponding to the recurrence $y_k = -\sum_{i=1}^{n_a} a_i(\rho) y_{k-i}$.

\subsection{Convex Reparameterization of $K(\rho)$ for Stability}
As a step towards an unconstrained reparameterization, this subsection reparameterizes all possible $P,K(\rho)$ for which \eqref{eq:contraction_F_GK} holds in terms of a convex set of variables $W$ and $M(\rho)$, turning constraint \eqref{eq:contraction_F_GK} that is non-convex in $P,K(\rho)$ into a convex one in $W,M(\rho)$. Condition \eqref{eq:contraction_F_GK} is rewritten as follows.
\begin{lemma}\label{lem:contraction_F_GK_riccati}
    For a given $K(\rho)$ as in \eqref{eq:maximum_ss_ABCD}, there exists a $P \succ 0$ satisfying stability condition \eqref{eq:quadratic_stability} if and only if $P$, $K(\rho)$ satisfy
    \begin{equation}
        \begin{aligned}     
        \label{eq:contraction_F_GK_riccati}
            P - F^\T P F + F^\T P G (G^\T P G)^{-1} G^\T P F \\ - H^\T(\rho) G^\T P G H(\rho) \succ 0 \quad \forall \rho \in \mathbb{P},
        \end{aligned}
    \end{equation}
    with 
    \begin{equation}
        \label{eq:H}
        H(\rho) = K(\rho) - (G^\T P G)^{-1} G^\T P F.
    \end{equation}
\end{lemma}
\begin{proof}
    The proof is based on completing the squares of the quadratic term ($K^\T(\rho) G^\T P G K(\rho)$) and linear term ($F^\T P G K(\rho)$) in \eqref{eq:contraction_F_GK}. First, note that $H(\rho)$ is well-defined: by the structure of $G$, $G^\T P G$ is the (1,1) entry of $P$, which is positive as $P \succ 0$, and thus $G^\T P G \in \mathbb{R}_{> 0}$ and its inverse are well-defined. Second, with $H(\rho)$ in \eqref{eq:H}, it holds that
    \begin{align}
        &\hspace{-5pt}-H^\T(\rho) G^\T P G H(\rho)  = K^\T(\rho) G^\T P F + F^\T P G K(\rho) \nonumber \\ 
        &\hspace{3pt}- K^\T(\rho) G^\T P G K(\rho) - F^\T P G (G^\T P G)^{-1} G^\T P F. \label{eq:completed_squares}
    \end{align}
        Expanding \eqref{eq:contraction_F_GK} gives
    \begin{equation}
    \begin{aligned}
        \label{eq:contraction_F_GK_expanded}
        P - F^\T P F + K^\T(\rho) G^\T P F + F^\T P G K(\rho) \\ - K^\T(\rho) G^\T P G K(\rho) \succ 0 \quad \forall \rho \in \mathbb{P}.
    \end{aligned}
    \end{equation}
    Substituting \eqref{eq:completed_squares} into \eqref{eq:contraction_F_GK_expanded} gives \eqref{eq:contraction_F_GK_riccati}. 
\end{proof}
While Lemma \ref{lem:contraction_F_GK_riccati} is just a reformulation of \eqref{eq:contraction_F_GK}, it is useful for characterizing all candidate $P$ that should be considered for \eqref{eq:contraction_F_GK}, as formalized by the following two corollaries.

\begin{corollary}
    \label{cor:riccati_inequality}
    Any $P \succ 0$ satisfying \eqref{eq:contraction_F_GK} has to satisfy
    \begin{equation}
        P - F^\T P F + F^\T P G  (G^\T P G)^{-1} G^\T P F \succ 0.
        \label{eq:riccati_inequality}
    \end{equation}
\end{corollary}

\begin{proof}
    $H^\T(\rho) G^\T P G H(\rho)$ is a quadratic form with $P \succ 0$ such that $H^\T(\rho) G^\T P G H(\rho) \succeq 0$, where non-strictness follows from the fact that $H(\rho) \in \mathbb{R}^{1 \times n_a}$ has rank 1. Thus \eqref{eq:contraction_F_GK_riccati} implies \eqref{eq:riccati_inequality} for any $P$ that satisfies \eqref{eq:contraction_F_GK}. 
\end{proof}

The above corollary illustrates that not any $P \succ 0$ can be a Lyapunov function because $F-GK(\rho)$ is structured: $P$ needs to additionally satisfy the Riccati inequality \eqref{eq:riccati_inequality}.
Given this specification of all $P$ that should be considered, all solutions $K(\rho),P$ that satisfy \eqref{eq:contraction_F_GK} can be reparameterized in terms of a convex set of matrices, as formalized next.
\begin{theorem}
    \label{th:convex_parameterization_contraction}
    The LPV-IO model \eqref{eq:LPV_IO} satisfies \eqref{eq:contraction_F_GK} with a $P \succ 0$ if and only if there exists an $M(\rho): \mathbb{P} \rightarrow \mathbb{R}^{1 \times n_a}$ with $M^\T(\rho) M(\rho) \prec I \ \forall \rho \in \mathbb{P}$ and $W \in \mathbb{S}^{n_a}_{\succ 0}$ such that
    \begin{align}
        P - F^\T P F + F^\T P G  (G^\T P G)^{-1} G^\T P F = W \label{eq:riccati_equality} \\
        K(\rho) = (G^\T P G)^{-1} G^\T P F + X_Q^{-1} M(\rho) X_W \label{eq:M_rho_to_K_rho}
    \end{align}
    where $X_W,X_Q$ are matrix factorizations given by $W = X_W^\T X_W$, $G^\T P G = X_Q^\T X_Q$.
\end{theorem}

\begin{proof}
\eqref{eq:contraction_F_GK} $\Rightarrow$ \eqref{eq:riccati_equality}-\eqref{eq:M_rho_to_K_rho}: If \eqref{eq:contraction_F_GK} is satisfied with a $P\succ 0$, then, by Corollary \ref{cor:riccati_inequality}, $P$ satisfies \eqref{eq:riccati_inequality}. Thus there exists a $W \succ 0$ such that \eqref{eq:riccati_equality} is satisfied. Since $W \succ 0$, it can be factorized as $W = X_W^\T X_W$ with $X_W$ full rank. Then \eqref{eq:contraction_F_GK_riccati} reads as
\begin{equation}
    X_W^\T X_W - H^\T(\rho) G^\T P G H(\rho) \succ 0 \quad \forall \rho \in \mathbb{P},
    \label{eq:standard_form_}
\end{equation}
with $H(\rho)$ as defined in \eqref{eq:H}. Since $P\succ0$, also $G^\T P G \succ 0$, such that it can be factorized as $G^\T P G = X_Q^\T X_Q$. 
Multiplying \eqref{eq:standard_form_} on the left with $X_W^{-\T}$ and on the right with $X_W^{-1}$, and substituting $G^\T P G = X_Q^\T X_Q$ results in
\begin{equation}
    I - X_W^{-\T} H^\T(\rho) X_Q^\T X_Q H(\rho) X_W^{-1} \succ 0 \quad \forall \rho \in \mathbb{P}.
    \label{eq:standard_form}
\end{equation}
Then \eqref{eq:standard_form} implies that there exists a $M(\rho) = X_Q H(\rho) X_W^{-1}$ that satisfies
\begin{equation}
    I \succ M^\T(\rho) M(\rho) \quad \forall \rho \in \mathbb{P}.
\end{equation}
Note that $M(\rho)$ is related to $H(\rho)$ as $H(\rho) = X_Q^{-1} M(\rho) X_W$ and thus by \eqref{eq:H} to $K(\rho)$ as
\begin{equation}
    K(\rho) = (G^\T P G)^{-1} G^\T P F + X_Q M(\rho) X_W^{-1},
\end{equation}
completing the only if part of the proof.

\eqref{eq:contraction_F_GK} $\Leftarrow$ \eqref{eq:riccati_equality}-\eqref{eq:M_rho_to_K_rho}: Given any $W\succ0$, by controllability of $F,G$, Riccati equation \eqref{eq:riccati_equality} has a unique positive definite solution $P \succ 0$ \cite{EmamiNaeini1982}. Then, given this $P\succ 0$ and $M(\rho)$, construct $K(\rho)$ as in \eqref{eq:M_rho_to_K_rho} and substitute it in \eqref{eq:contraction_F_GK} to obtain
\begin{equation}
\begin{aligned}
    &P - (F - G K(\rho))^\T P (F - G K(\rho)) = P - F^\T P F  \\
    &+ F^\T P G \left( (G^\T P G)^{-1} G^\T P F + X_Q^{-1} M(\rho) X_W \right)  \\
    &+ \left( F^\T P G (G^\T P G)^{-1} + X_W^\T M^\T(\rho) X_Q^{-\T} \right) G^\T P F  \\
    &- F^\T P G (G^\T P G)^{-1} G^\T P G (G^\T P G)^{-1} G^\T P F  \\ 
    & - X_W^\T M^\T(\rho) X_Q^{-\T} G^\T P F  - F^\T P G X_Q^{-1} M(\rho) X_W \\
    &- X_W^\T M^\T(\rho) X_Q^{-\T} G^\T P G X_Q^{-1} M(\rho) X_W,
\end{aligned}
\end{equation}
which can be simplified to
\begin{equation}
\begin{aligned}
    & P - F^\T P F + F^\T P G (G^\T P G)^{-1} G^\T P F \\
    &- X_W^\T M^\T(\rho) X_Q^{-\T} G^\T P G X_Q^{-1} M(\rho) X_W. \\
\end{aligned}
\end{equation}
Using \eqref{eq:riccati_equality} and $G^\T P G = X_Q^\T X_Q$, this is equivalent to
\begin{align}
    W \! - \! X_W^\T M^\T(\rho) M(\rho) X_W \!= \!X_W^\T (I \! - \! M^\T(\rho) M(\rho) ) X_W \succ 0,
    \nonumber
\end{align}
where positive-definiteness follows by $M^\T(\rho) M(\rho) \prec I \ \forall \rho \in \mathbb{P}$, i.e., the constructed $K(\rho),P$ satisfy \eqref{eq:quadratic_stability}.
\end{proof}

Theorem \ref{th:convex_parameterization_contraction} can be interpreted as a small-gain result, stating that the spectral norm of $M(\rho)$ should be smaller than 1 for all $\rho$, i.e., $M^\T(\rho) M(\rho) \prec I$ or equivalently $\norm{M(\rho)}_2^2 < 1 \ \forall \rho \in \mathbb{P}$. Theorem \ref{th:convex_parameterization_contraction} then states that all coefficient functions $K(\rho)$ that result in a quadratically stable LPV-IO model can be generated from transformed coefficient functions $M(\rho)$ constrained to the unit ball, and a rotation, scaling and translation of this unit ball based on $P$ through \eqref{eq:M_rho_to_K_rho}. The set of allowable $P$ is described by the image of the positive-definite cone $W \succ 0$ under the Riccati equation \eqref{eq:riccati_equality}. Last, since $\{ W \ | \ W\succ0\}$ and $\{ M(\rho) \ | \ M^\T(\rho) M(\rho) \prec I\}$ are convex sets, Theorem \ref{th:convex_parameterization_contraction} illustrates that the set of all quadratically stable LPV-IO models is convex in $W$ and the transformed coefficient functions $M(\rho)$.

\subsection{Unconstrained Reparameterization of $K(\rho)$ for Stability}
To obtain an unconstrained parameterization, the convex constraints on variables $W,M(\rho)$ of Theorem \ref{th:convex_parameterization_contraction} are reparameterized directly in terms of unconstrained variables $X_W$ related to $W$ and $X_M(\rho),Z_M(\rho)$ related to $M(\rho)$. 

Specifically, the convex set of norm-bounded matrices can be generated from free matrices through a Cayley transformation, as is defined next.

\begin{lemma}
    \label{lem:cayley}
    Given $M \in \mathbb{R}^{n \times m}$ with $n \geq m$. Then $M^\T M \prec I$ if and only if there exist $X_M, Y_M \in \mathbb{R}^{m \times m}$, $Z_M \in \mathbb{R}^{n - m \times m}$ with $X_M$ full rank such that
    \begin{equation}
        M = 
        \begin{bmatrix}
            (I-N)(I+N)^{-1} \\
            -2Z_M (I + N)^{-1}
        \end{bmatrix}
    \end{equation}
    with $N = X_M^\T X_M + Y_M - Y_M^\T + Z_M^\T Z_M$.
\end{lemma}
\begin{remark}
    \inputnewtext{The rank condition on $X_M$ is not restrictive: without loss of generality, $X_M$ can be chosen upper triangular with its diagonal equal to $e^d$ with $d \in \mathbb{R}^{m}$ a free vector.}
\end{remark}
For a proof, see \cite[Lemma 1]{Verhoek_CDC_2023}. Lemma \ref{lem:cayley} states that any matrix that is norm-bounded by 1 can be formed through a Cayley transformation of a passive matrix $N$, that is formed as a positive definite term $X^\T X + Z^\T Z$ and a skew-symmetric term $Y - Y^\T$. Given Lemma \ref{lem:cayley}, the convex constraint $M^\T(\rho) M(\rho) \prec I $ of Theorem \ref{th:convex_parameterization_contraction} is satisfied if and only if there exist unconstrained matrix functions related to $M(\rho)$ through this Cayley transformation, as shown next.
\begin{theorem}
    \label{th:direct_parameterization_contraction}
    The LPV-IO model \eqref{eq:LPV_IO} satisfies \eqref{eq:contraction_F_GK} with $P\succ0$ if and only if there exist $X_M(\rho): \mathbb{P} \rightarrow \mathbb{R}$, $Z_M(\rho): \mathbb{P} \rightarrow \mathbb{R}^{n_a - 1 \times 1}$, $X_W \in \mathbb{R}^{n_a \times n_a }$ with $X_M(\rho) \neq 0 \ \forall \rho \in \mathbb{P}$ and $X_W$ full rank such that
    \begin{align}
        W &= X_W^\T X_W \label{eq:X_W_to_W} \\
        M(\rho) &= \begin{bmatrix} 
        \textnormal{Cayley}(N(\rho)) \\
        -2Z(\rho)(I+N(\rho))^{-1}
        \end{bmatrix}^\T
        \label{eq:N_Z_to_M}
    \end{align}
    with $N(\rho) = X_M^\T(\rho) X_M(\rho) + Z_M^\T(\rho) Z_M(\rho)$ and $W$, $M(\rho)$ satisfying \eqref{eq:riccati_equality}-\eqref{eq:M_rho_to_K_rho}.
\end{theorem}
\begin{proof}
    A matrix $W$ is positive definite if and only if it can be factorized as $W = X_W^\T X_W$ with $X_W$ square and full rank. Additionally, by Lemma \ref{lem:cayley} with $m=1$, a matrix function $M(\rho): \mathbb{P} \rightarrow \mathbb{R}^{1 \times n_a}$ is norm bounded $\forall \rho \in \mathbb{P}$, i.e., $M^\T(\rho) M(\rho) \prec I \ \forall \rho \in \mathbb{P}$ if and only if there exists $X_M(\rho)$ and $Z_M(\rho)$ as stated, since $Y - Y^\T = 0$ for $n_u = 1$. The rest of the proof follows by Theorem \ref{th:convex_parameterization_contraction}.
\end{proof}

In the above theorem, $X_M(\rho)$ and $Z_M(\rho)$ can be any function of $\rho$ and the transformations \eqref{eq:X_W_to_W},\eqref{eq:N_Z_to_M},\eqref{eq:riccati_equality},\eqref{eq:M_rho_to_K_rho} guarantee that the resulting LPV-IO model in terms of $K(\rho)$ is stable with $P \succ 0$ a Lyapunov function for $F-GK(\rho)$ as in \eqref{eq:contraction_F_GK}. Consequently, an unconstrained parameterization of all quadratically stable DT-LPV-IO models is obtained.

\begin{remark}
    Note that for a non-varying $\rho$, Theorem \ref{th:direct_parameterization_contraction} gives an unconstrained reparameterization of all stable linear time-invariant DT transfer functions.
\end{remark}

\begin{figure}
    \centering
    \includegraphics[width=\columnwidth]{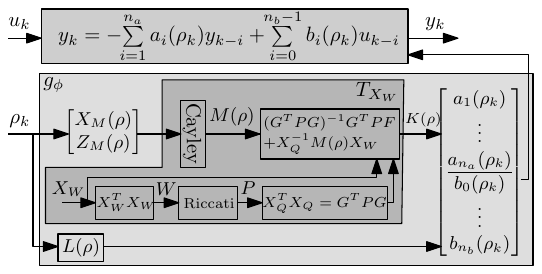}
    \caption{Graphic representation of the stable DT-LPV-IO model class. For any value of $X_W$, and for any parameterized functions $X_M(\rho),Z_M(\rho),L(\rho)$, the coefficient functions $a_i(\rho),b_i(\rho)$ result in a stable DT-LPV-IO model.}
    \label{fig:stable_LPV_IO_model_class}
\end{figure}

Zooming out to model class \eqref{eq:LPV_IO} with coefficient functions $a_i(\rho),b_i(\rho)$ described by $g_\phi$ as in \eqref{eq:coefficient_functions}, Theorem \ref{th:direct_parameterization_contraction} states that if a quadratically stable DT-LPV-IO model is desired, it can be represented in an unconstrained way by a $g_\phi$ that consists two elements: matrix functions $X_M(\rho), Z_M(\rho), L(\rho)$ that describe the coefficient functions in a transformed space dependent on one set of model parameters $\phi_1$, and a transformation $T_{X_W}$ that maps $X(\rho), Z_M(\rho)$ to $K(\rho)$
dependent on a second set of model parameters $\phi_2 = \vecc(X_W)$. Matrix functions $X(\rho), Z_M(\rho), L(\rho)$ can be any function parameterized by $\phi_1$, e.g., an affine map, polynomial expansion or neural network, see also Section \ref{sec:problem_formulation}, and $T_{X_W}$ ensures that the DT-LPV-IO model with the resulting $a_i(\rho),b_i(\rho)$ is quadratically stable. The quadratically stable DT-LPV-IO model is visualized in Fig. \ref{fig:stable_LPV_IO_model_class}.

%% file: Sections/Simulation_example.tex
\section{Application to System Identification}
\label{sec:simulation_example}
In this section, the unconstrained parameterization of quadratically stable DT-LPV-IO models of Theorem \ref{th:direct_parameterization_contraction} is applied in a system identification setup \footnote{The code used to generate the example can be found at \protect\url{https://gitlab.tue.nl/kon/stable-lpv-io-estimation}}.

\subsection{LPV Output-Error System Identification Setup}
The considered data-generating system $\mathcal{G}: u_k,\rho_k \rightarrow \tilde{y}_k$, with $u_k,\rho_k,\tilde{y}_k\in\mathbb{R}$ is given by the LPV-IO representation
\begin{equation}
\begin{aligned}
    u_k &= m \delta^2 \tilde{y}_k + c \delta \tilde{y}_k + k(\rho_k) \tilde{y}_k\\
    y_k &= \tilde{y}_k + v_k,
    \label{eq:plant}
\end{aligned}
\end{equation}
with $y_k \in \mathbb{R}$ the measurement of the true output $\tilde{y}_k$ corrupted by i.i.d. white noise $v$ with $\mathbb{E}(v^2) = \sigma_v^2$, resulting in an LPV output-error (OE) identification setup\footnote{More generic noise model structures can easily be incorporated, but for ease of notation, an LPV-OE setting is considered.}. Here $\delta = (1-q^{-1})/T_s$ is the backward difference operator, $q$ is the forward-time shift, e.g., $qy_k = y_{k+1}$, and $T_s$ is the sampling time. Consequently, $\mathcal{G}$ can be recognized as the Euler discretization of a mass-damper-spring system with parameter-varying stiffness $k(\rho)$ and fixed mass and damping $m,d$. By expanding $\delta$, $\mathcal{G}$ can be written as \eqref{eq:LPV_IO} with $n_a=2,n_b=1$. \inputnewtext{Lastly, an LMI check shows that $\mathcal{G}$ is QS, such that it can be represented by the stable DT-LPV-IO model.}

\begin{figure}[t]
    \centering
    \includegraphics{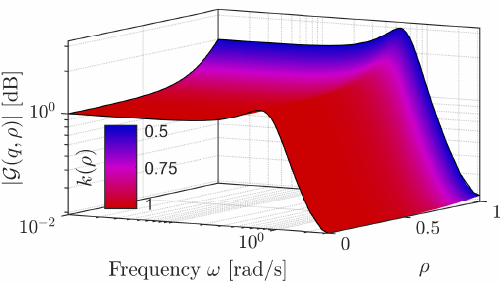}
    \caption{Bode plot of frozen dynamics of $\mathcal{G}(\delta,\rho)$, displaying a resonance with scheduling-dependent damping.}
    \label{fig:LPV_bode}
\end{figure}

In this simulation example, $\rho \in \mathbb{P} = [0,1]$, $T_s = 1$ and the true coefficient functions are given by
\begin{equation}
    \begin{aligned}
        m  = 1 & & d  = 0.1 & & k(\rho) = 1 - \frac{1}{1 + e^{-7 \rho + 7}}
    \end{aligned}
\end{equation}
where $k(\rho)$ represents, e.g., spring softening as a function of temperature. These coefficient functions result in frozen LTI behaviour of $\mathcal{G}$ visualized in Fig. \ref{fig:LPV_bode}.

 A dataset $\mathcal{D} = \{u_k,y_k,\rho_k\}_{k=1}^N$ of length $N=1000$ samples is generated by $\mathcal{G}$ with $u_k = \sum_{i=1}^{10} \sin(2 \pi f_i k/T_s)$ a multisine with $f_i$ linearly spaced between $0.01$ and $0.1$ Hz, and $\rho_k = 1-k N^{-1}$ a linear scheduling trajectory. The noise variance is set as $\sigma_v^2 = 0.1$ for signal-to-noise ratio $10 \log_{10}(\frac{\norm{y}^2}{\norm{v}^2}) = 19.5$ dB.

\subsection{Model Parameterization and Identification Criterion}
A stable DT-LPV-IO model with $n_a = 2, n_b = 1$ is chosen as a model for $\mathcal{G}$, i.e., the model has the same order as $\mathcal{G}$. Then any parameterization for $L(\rho),X_M(\rho_k),Z_M(\rho_k)$ can be considered, and the model can be optimized using prediction-error minimization based on unconstrained gradient-based optimization \cite{Zhao2012a,KonCDC2023}.

Specifically, in this paper, the transformed coefficient functions $X_M(\rho),Z_M(\rho),L(\rho)$ are parameterized as
\begin{equation}
        \begin{bmatrix}
            X_M \hspace{-2pt} & \hspace{-2pt} Z_M \hspace{-2pt} & \hspace{-2pt} L
        \end{bmatrix}\! (\rho)
        = E_2 \sigma( E_1 \sigma( E_0 \rho +  c_0))  + c_1) + c_2,
\end{equation}
with $\sigma = \tanh$, $E_0 \in \mathbb{R}^{5 \times 1}, E_1 \in \mathbb{R}^{5 \times 5}, E_2 \in \mathbb{R}^{3 \times 5}, c_0 \in \mathbb{R}^5, c_1 \in \mathbb{R}^5, c_2 \in \mathbb{R}^3$, i.e., a neural network with $2$ hidden layers of 5 nodes each and 3 outputs since $L(\rho_k),X_M(\rho_k),Z_M(\rho_k)\in\mathbb{R}$ for $n_a = 2, n_b = 1$. Consequently, the model parameters are $\phi =\vecc(E_0,c_0,\ldots E_2,c_2,X_W) \in \mathbb{R}^{n_\phi}$ with $X_W \in \mathbb{R}^{2\times2}$ upper triangular such that $n_\phi = 58 + 3$.

In the above OE setting with noiseless $\rho_k$, the model parameters $\phi$ are found by minimizing the $\ell_2$ loss of the prediction error $V_N(\phi)$ as $\phi^* = \arg \min_\phi V_N(\phi)$ with 
\begin{equation}
    V_N(\phi) = \sqrt{\frac{1}{N} \sum_{k=1}^N (y_k - \hat{y}_{k,\phi})^2},
    \label{eq:V_OE}
\end{equation}
where $\hat{y}_{k,\phi}$ is the simulated model response.

Criterion \eqref{eq:V_OE} is optimized using the Levenberg-Marquardt optimization algorithm with finite differencing for Jacobian estimation, resulting in a training time of 37 seconds on an Z-book G5 using a Intel Core i7-8750H CPU. A significant speedup for longer datasets can be achieved by employing analytic Jacobian expressions, see \cite{KonCDC2023}.

\subsection{Identification with Stability Guarantees} 
\begin{figure}[t]
    \centering
    \includegraphics{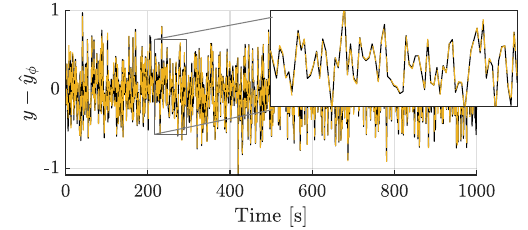}
    \includegraphics{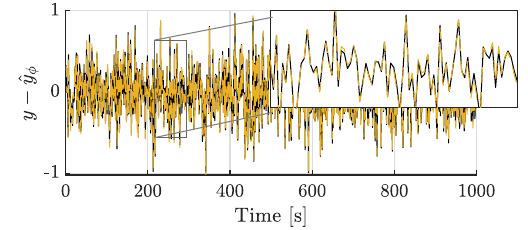}
    \caption{Train (top) and validation (bottom) prediction residuals $y-\hat{y}_\phi$ (\protect \drawlinelegend{morange}) and noise realization $y - \tilde{y}$ (\protect \drawlinelegend{black}). The prediction residuals coincide with the noise, i.e., the LPV-IO model \eqref{eq:LPV_IO} with neural network coefficient functions is able to learn all dynamics while simultaneously guaranteeing stability.}
    \label{fig:residuals}
\end{figure}

Fig. \ref{fig:residuals} show the residuals $y-\hat{y}_\phi$ after optimization for the training dataset $\mathcal{D}$ and a similar but different validation dataset. For the training dataset, the estimated parameter vector $\phi^*$ achieves $V_N(\phi^*) = 0.313$, corresponding to the noise level with $\sigma_v = 0.316$
illustrating that the only contribution to $V_N(\phi^*)$ is noise that cannot be predicted. Similarly, for the validation dataset, $V_N(\phi^*) = 0.320$, indicating that the model can generalize well.

\subsection{Visualization of Stability Sets}
\begin{figure}[t]
    \centering
    \includegraphics{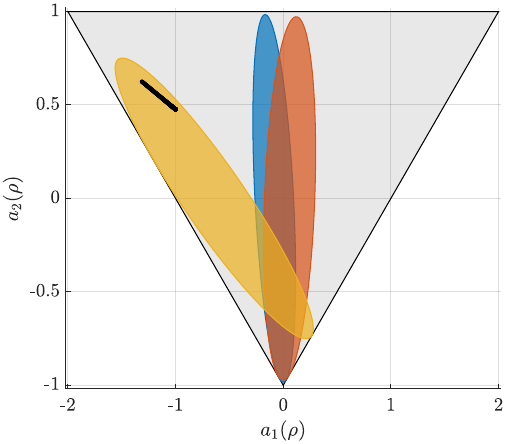}
    \caption{Coefficient set $K_P = \{K(\rho) = \begin{bmatrix} a_1(\rho),a_2(\rho) \end{bmatrix} \ | \ P \succ 0, (F-GK(\rho))^\T P (F-GK(\rho))- P \prec 0 \ \forall \rho \in \mathbb{P} \}$, i.e., the set of all possible values $K(\rho)$ such that LPV IO-model \eqref{eq:maximum_ss_short} is stable with Lyapunov certificate $P$ at iteration 1 (\protect\drawrectanglelegend{mblue}{40}), 10 (\protect\drawrectanglelegend{mred}{40}), and 100 (\protect\drawrectanglelegend{morange}{40}) of the optimization. Lyapunov certificate $P$ is optimized such that the true coefficient functions (\protect\drawrectanglelegendsmall{black}{100}) are contained within the stable coefficient set. Optimizing $P$ thus corresponds to rotating, scaling and translating this set. All stable coefficient sets are included in the stability triangle that describes the stable coefficients for the LTI case (\protect\drawrectanglelegend{black}{10}). }
    \label{fig:stability_sets}
\end{figure}

In this section, the evolution of the coefficient set that can be represented by the DT-LPV-IO model during the iterations of the optimization is visualized, i.e., the set $K_P$ in which $a_i(\rho),b_i(\rho)$ can take values $\forall \rho$ by construction, see Theorem \ref{th:direct_parameterization_contraction}. Specifically, given $X_W$ during optimization, all coefficients $K(\rho) = \begin{bmatrix} a_1(\rho) & a_2(\rho) \end{bmatrix}$ corresponding to this $X_W$ can be constructed using \eqref{eq:riccati_equality},\eqref{eq:M_rho_to_K_rho},\eqref{eq:X_W_to_W},\eqref{eq:N_Z_to_M}, resulting in the coefficient sets $K_P$ of Fig. \ref{fig:stability_sets}. The following observations are made.
\begin{itemize}
    \item Mapping \eqref{eq:M_rho_to_K_rho} corresponds to scaling, rotating and translating the unit ball $\norm{M(\rho)}_2^2 < 1$, resulting in the ellipsoidal shape of $K_P$. Thus, $P$ can be thought of as describing all possible rotations, translations and scalings of $K(\rho)$ such that LPV-IO model \eqref{eq:LPV_IO} is stable.
    \item Graphically, each $K_P$ visualizes a set in which the function $K(\rho)$ can generate outputs for the LPV-IO model to be stable. Thus, the true coefficient functions necessarily have to be fully contained in a $K_P$ for $K(\rho)$ to be able to describe them. Thus, optimizing $X_W$ is equivalent to transforming the ellipsoid such that it encapsulates the true coefficient functions.
    \item Each $K_P$ is necessarily contained within the triangle of stable LTI transfer function coefficients. Moreover, the full triangle can be filled by the union of all stable coefficient sets.
    \item For an LPV-IO model to be quadratically stable, there needs to exist a $K_P$ corresponding to some $P$ that fully encapsulates the true coefficient functions. Fig. \ref{fig:stability_sets} thus provides a graphical tool for accessing stability properties of an LPV-IO model.
\end{itemize}

%% file: Sections/Conclusion.tex
\section{Conclusion}
\label{sec:conclusion}
In this paper, the class of all quadratically stable DT-LPV-IO models is reparameterized in terms of unconstrained model parameters. This unconstrained parameterization is achieved through reparameterizing the quadratic stability condition in a necessary and sufficient fashion through a Riccati equation and a Cayley transformation. It allows for using arbitrary dependency of the scheduling coefficients on the scheduling signal $\rho$, e.g., a polynomial or NN.

The resulting stable DT-LPV-IO model class enables system identification with a priori stability guarantees on the identified model in the presence of modeling errors and measurement noise. Since it does not require enforcing an LMI condition during optimization, it can be optimized using standard unconstrained optimization routines, significantly decreasing the computational complexity. Additionally, the unconstrained stable model class allows for, i.a., sampling of stable DT-LPV-IO systems. 

%% file: ms.bbl
\begin{thebibliography}{10}
\providecommand{\url}[1]{#1}
\csname url@samestyle\endcsname
\providecommand{\newblock}{\relax}
\providecommand{\bibinfo}[2]{#2}
\providecommand{\BIBentrySTDinterwordspacing}{\spaceskip=0pt\relax}
\providecommand{\BIBentryALTinterwordstretchfactor}{4}
\providecommand{\BIBentryALTinterwordspacing}{\spaceskip=\fontdimen2\font plus
\BIBentryALTinterwordstretchfactor\fontdimen3\font minus
  \fontdimen4\font\relax}
\providecommand{\BIBforeignlanguage}[2]{{%
\expandafter\ifx\csname l@#1\endcsname\relax
\typeout{** WARNING: IEEEtran.bst: No hyphenation pattern has been}%
\typeout{** loaded for the language `#1'. Using the pattern for}%
\typeout{** the default language instead.}%
\else
\language=\csname l@#1\endcsname
\fi
#2}}
\providecommand{\BIBdecl}{\relax}
\BIBdecl
\renewcommand{\BIBentryALTinterwordstretchfactor}{4}

\bibitem{Schoukens2019}
J.~Schoukens and L.~Ljung, ``{Nonlinear System Identification: A User-Oriented
  Road Map},'' \emph{IEEE Control Syst. Mag.}, vol. 39 (6), 2019.

\bibitem{Toth2010}
R.~T{\'{o}}th, \emph{{Modeling and identification of linear parameter-varying
  systems}}, ser. Lecture notes in control and information sciences vol.
  403.\hskip 1em plus 0.5em minus 0.4em\relax Springer, 2010.

\bibitem{Leith2000a}
D.~J. Leith and W.~E. Leithead, ``{On formulating nonlinear dynamics in LPV
  form},'' in \emph{Proc. IEEE Conf. Decis. Control}, 2000.

\bibitem{Bamieh2002}
B.~Bamieh and L.~Giarr{\'{e}}, ``{Identification of linear parameter varying
  models},'' \emph{Int. J. Robust Nonlinear Control}, vol. 12 (9), 2002.

\bibitem{Laurain2010}
V.~Laurain, M.~Gilson, R.~T{\'{o}}th, and H.~Garnier, ``{Refined instrumental
  variable methods for identification of LPV Box-Jenkins models},''
  \emph{Automatica}, vol. 46 (6), pp. 959--967, 2010.

\bibitem{Zhao2012a}
Y.~Zhao, B.~Huang, H.~Su, and J.~Chu, ``{Prediction error method for
  identification of LPV models},'' \emph{J. Process Control}, vol. 22 (1),
  2012.

\bibitem{Felici2007}
F.~Felici, J.~W. van Wingerden, and M.~Verhaegen, ``{Subspace identification of
  MIMO LPV systems using a periodic scheduling sequence},'' \emph{Automatica},
  vol.~43, no.~10, pp. 1684--1697, 2007.

\bibitem{Cox2018a}
P.~B. Cox, R.~T{\'{o}}th, and M.~Petreczky, ``{Towards efficient maximum
  likelihood estimation of LPV-SS models},'' \emph{Automatica}, vol.~97, 2018.

\bibitem{Lacy2003}
S.~L. Lacy and D.~S. Bernstein, ``{Subspace identification with guaranteed
  stability using constrained optimization},'' \emph{IEEE Trans. Automat.
  Contr.}, vol. 48 (7), pp. 1259--1263, 2003.

\bibitem{Haene2006}
T.~D'haene, R.~Pintelon, and G.~Vandersteen, ``{An Iterative Method to
  Stabilize a Transfer Function in the s- and z-domains},'' \emph{IEEE Trans.
  Instrum. Meas.}, vol. 55 (4), pp. 1192--1196, 2006.

\bibitem{Apkarian1995}
P.~Apkarian and P.~Gahinet, ``{A convex characterization of gain-scheduled
  $H_\infty$ controllers},'' \emph{IEEE Trans. Automat. Contr.}, vol. 40 (5),
  pp. 853--864, 1995.

\bibitem{Scherer2001}
C.~W. Scherer, ``{LPV control and full block multipliers},'' \emph{Automatica},
  vol. 37 (3), pp. 361--375, 2001.

\bibitem{Revay2023_REN}
M.~Revay, R.~Wang, and I.~R. Manchester, ``{Recurrent Equilibrium Networks:
  Flexible Dynamic Models with Guaranteed Stability and Robustness},''
  \emph{IEEE Trans. Automat. Contr.}, 2023.

\bibitem{Martinelli2023}
D.~Martinelli, C.~L. Galimberti, I.~R. Manchester, L.~Furieri, and
  G.~Ferrari-Trecate, ``{Unconstrained Parametrization of Dissipative and
  Contracting Neural Ordinary Differential Equations},'' in \emph{Proc. 62nd
  IEEE Conf. Decis. Control}, 2023.

\bibitem{Verhoek_CDC_2023}
C.~Verhoek, R.~Wang, and R.~T{\'{o}}th, ``{Learning Stable and Robust Linear
  Parameter-Varying State-Space Models},'' in \emph{Proc. 62nd IEEE Conf.
  Decis. Control}, 2023.

\bibitem{Cerone2012}
V.~Cerone, D.~Piga, D.~Regruto, and R.~T{\'{o}}th, ``{Input-output LPV model
  identification with guaranteed quadratic stability},'' in \emph{Proc. 16th
  IFAC Symp. Syst. Identif.}, 2012, pp. 1767--1772.

\bibitem{Cerone2012a}
V.~Cerone, D.~Piga, D.~Regruto, and R.~Toth, ``{Fixed order LPV controller
  design for LPV models in input-output form},'' \emph{Proc. IEEE Conf. Decis.
  Control}, pp. 6297--6302, 2012.

\bibitem{Wollnack2016}
S.~Wollnack and H.~Werner, ``{LPV-IO controller design: An LMI approach},'' in
  \emph{Am. Control Conf.}, 2016, pp. 4617--4622.

\bibitem{Wollnack2017}
S.~Wollnack, H.~S. Abbas, R.~T{\'{o}}th, and H.~Werner, ``{Fixed-structure
  LPV-IO controllers: An implicit representation based approach},''
  \emph{Automatica}, vol.~83, pp. 282--289, 2017.

\bibitem{Henrion2003}
D.~Henrion, D.~Peaucelle, D.~Arzelier, and M.~Sebek, ``{Ellipsoidal
  approximation of the stability domain of a polynomial},'' \emph{IEEE Trans.
  Automat. Contr.}, vol. 48 (12), pp. 2255--2259, 2003.

\bibitem{Gilbert2007}
W.~Gilbert, D.~Henrion, J.~Bernussou, and D.~Boyer, ``{Polynomial LPV synthesis
  applied to turbofan engines},'' in \emph{17th IFAC Symp. Autom. Control
  Aerosp.}, 2007, pp. 645--650.

\bibitem{Revay2021}
M.~Revay, R.~Wang, and I.~R. Manchester, ``{A Convex Parameterization of Robust
  Recurrent Neural Networks},'' in \emph{Am. Control Conf.}, 2021.

\bibitem{Revay2022}
M.~Revay, R.~Wang, and I.~R. Manchester, ``{Recurrent Equilibrium Networks:
  Unconstrained Learning of Stable and Robust Dynamical Models},'' in
  \emph{60th IEEE Conf. Decis. Control}, 2021.

\bibitem{Previdi2003}
F.~Previdi and M.~Lovera, ``{Identification of a class of non-linear
  parametrically varying models},'' \emph{Int. J. Adapt. Control Signal
  Process.}, vol. 17 (1), pp. 33--50, 2003.

\bibitem{KonCDC2023}
J.~Kon, J.~van~de Wijdeven, D.~Bruijnen, R.~T{\'{o}}th, M.~Heertjes, and
  T.~Oomen, ``{Direct Learning for Parameter-Varying Feedforward Control: A
  Neural-Network Approach},'' in \emph{62nd IEEE Conf. Decis. Control}, 2023.

\bibitem{Toth2012b}
R.~Toth, H.~S. Abbas, and H.~Werner, ``{On the state-space realization of LPV
  input-output models: Practical approaches},'' \emph{IEEE Trans. Control Syst.
  Technol.}, vol. 20 (1), 2012.

\bibitem{Toth2013}
R.~T{\'{o}}th, ``{Maximum LPV-SS realization in a static form},'' Eindhoven
  University of Technology, Tech. Rep., 2013.

\bibitem{EmamiNaeini1982}
A.~Emami-Naeini and G.~Franklin, ``{Deadbeat control and tracking of
  discrete-time systems},'' \emph{IEEE Trans. Automat. Contr.}, vol.~27, 1982.

\end{thebibliography}
